\documentclass[showpacs,twocolumn,groupedaddress,pra,aps]{revtex4-2}

\usepackage{amsmath}
\usepackage{amssymb}
\usepackage{maybemath}
\usepackage{slashed}
\usepackage{xcolor}
\usepackage{graphicx}
\usepackage{bm}
\usepackage[caption=false]{subfig}
\usepackage{braket}
\usepackage{mathrsfs}
\usepackage{tikz}

\usepackage{changes}
\definechangesauthor[name={Natalia}, color=magenta]{nat}

\usetikzlibrary{decorations.pathmorphing}
\usetikzlibrary{shapes}
\usetikzlibrary{shapes.geometric}

\tikzset{
    magnetic/.style={
        fill,
        shape border rotate=-90,
        isosceles triangle,
        isosceles triangle apex angle=60,
        node distance=1,
        minimum height=.1
    }
}

\tikzset{
    othermagnetic/.style={
        fill,
        shape border rotate=90,
        isosceles triangle,
        isosceles triangle apex angle=60,
        node distance=1,
        minimum height=.1
    }
}

\tikzset{
    leftmagnetic/.style={
        fill,
        shape border rotate=0,
        isosceles triangle,
        isosceles triangle apex angle=60,
        node distance=1,
        minimum height=.1
    }
}

\tikzset{
    rightmagnetic/.style={
        fill,
        shape border rotate=180,
        isosceles triangle,
        isosceles triangle apex angle=60,
        node distance=1,
        minimum height=.1
    }
}

\pgfdeclaredecoration{complete sines}{initial}
{
    \state{initial}[
        width=+0pt,
        next state=upsine,
        persistent precomputation={\pgfmathsetmacro\matchinglength{
            \pgfdecoratedinputsegmentlength / int(\pgfdecoratedinputsegmentlength/\pgfdecorationsegmentlength)}
            \setlength{\pgfdecorationsegmentlength}{\matchinglength pt}
        }] {}
    \state{upsine}[width=\pgfdecorationsegmentlength,next state=downsine]{
        \pgfpathsine{\pgfpoint{0.25\pgfdecorationsegmentlength}{0.5\pgfdecorationsegmentamplitude}}
        \pgfpathcosine{\pgfpoint{0.25\pgfdecorationsegmentlength}{-0.5\pgfdecorationsegmentamplitude}}
    }
    \state{downsine}[width=\pgfdecorationsegmentlength,next state=upsine]{
        \pgfpathsine{\pgfpoint{0.25\pgfdecorationsegmentlength}{-0.5\pgfdecorationsegmentamplitude}}
        \pgfpathcosine{\pgfpoint{0.25\pgfdecorationsegmentlength}{0.5\pgfdecorationsegmentamplitude}}
}
    \state{final}{}
}

\usepackage{dcolumn}
\newcolumntype{.}{D{.}{.}{-1}}

\usepackage[%
  colorlinks=true,
  urlcolor=blue,
  linkcolor=blue,
  citecolor=blue
]{hyperref}

\makeatletter

\newcommand{\Rmnum}[1]{\expandafter\@slowromancap\romannumeral #1@}
\makeatother

\newcommand{\balpha}{\mbox{\boldmath $\alpha$} }

\begin{document}

\newcommand{\addrMPIK}{Max Planck Institute for Nuclear Physics, Saupfercheckweg 1, 69117 Heidelberg, Germany}

\title{Radiative and photon-exchange corrections to New Physics contributions to energy levels in few-electron ions} 

\author{V. Debierre}
\email[]{vincent.debierre@mpi-hd.mpg.de}
\author{N.~S. Oreshkina}
\email[]{natalia.oreshkina@mpi-hd.mpg.de}
\affiliation{\addrMPIK}

\begin{abstract}
The influence of hypothetical new interactions beyond the Standard Model on atomic spectra has attracted recent interest. In the present work, interelectronic photon-exchange corrections and radiative quantum electrodynamic corrections to the hypothetical contribution to the energy levels of few-electron ions from a new interaction are calculated. The $1s$, $2s$ and $2p_{1/2}$ ground states of H-like, Li-like and B-like ions are considered, as motivated by proposals to use isotope shift spectroscopy of few-electron ions in order to set stringent constraints on hypothetical new interactions. It is shown that, for light Li-like and B-like ions, photon-exchange corrections are comparable to or even larger, by up to several orders of magnitude, than the leading one-electron contribution from the new interaction, when the latter is mediated by heavy bosons.
\end{abstract}


\maketitle

\section{Introduction} \label{sec:Intro}

Precision spectroscopy of one- and few-electron ions is a powerful tool for testing fundamental physical theories~\cite{Weinberg1,ProtonRadiusSolved,Micke,LiLikeBi,LitaChantler,CalciumIS,ShabaevReview,ConfReview,IndelicatoReview}. It has also allowed the most precise determination of the electron mass~\cite{Sturm14,ExtractJacek} at the time. Moreover, proposals have been put forward to use comparisons of measurements and calculations on $g$ factors of highly-charged ions to obtain an improved determination of the fine-structure constant $\alpha$ of electrodynamics~\cite{WDiffOld,GFactorAlpha,HalilReduced}. The high-precision regime in which both experimental and theoretical efforts operate also motivates proposals~\cite{FifthForceG,HCIClockApp,RehbehnCoronal} to use highly charged ions to search for physics beyond the Standard Model (SM), also known simply as New Physics (NP). Indeed, NP would bring small contributions to precisely measured and calculated spectroscopic quantities~\cite{FifthForce,PossibleForces}, such as energy levels and $g$ factors. This is the premise for the search for NP at the precision frontier~\cite{LowEFrontier,CoulombHidden,SearchNPAtMol,HCIClockApp}: when experiment and theory agree at a certain level of accuracy, NP can be constrained at that same level, which in some cases can be competitive~\cite{LowEFrontier,CoulombHidden,FifthForce,FifthForceG,NonLinXPYb,LinXPCa} with constrained obtained from high-energy physics and cosmology. Conversely, a disagreement between experiment and theory might be a sign of NP.\\

The purpose of this work is to calculate quantum electrodynamic (QED) corrections, coming from interelectronic interactions (IEI) as well as radiative processes, to the hypothetical NP contribution to ionic energy levels. It is naturally expected that, if they exist, NP contributions to energy levels would be very small. Hence, it could seem that considering QED corrections to these hypothetical contributions is a purely academic exercise of little practical relevance, since these corrections can typically be expected to be even smaller than the leading NP contribution. However, for highly localised potentials, the second-order contributions can be comparable to, or even up to some orders of magnitude larger than the leading contribution from that potential~\cite{FNSVP,GFactorB}. In our case, the leading contribution is the one-electron hypothetical NP correction, generated by a Yukawa potential. Our results do show that interelectronic interactions and, to a lesser extent, radiative corrections, bring contributions which can compete with those of the leading one-electron contribution. Our analysis thus provides a more secure footing to the search for NP with few-electron ions. Moreover, we note that Li-like and B-like ions are more accessible to spectroscopy than H-like ions, since their low-lying transitions are of lower frequency, making the former promising candidates for experiments (see Ref.~\cite{RehbehnCoronal}).\\

In Sec.~\ref{sec:Frontier}, we briefly review the main guiding lines of the search for NP at the so-called precision frontier. In Sec.~\ref{sec:Leading}, we give the leading contribution from a hypothetical fifth force to the energy levels of few-electron ions. In Sec.~\ref{sec:Energy}, we calculate IEI and radiative QED corrections to that FF contribution for the energy levels. Numerical results are given in Sec.~\ref{sec:Res}. Finally, we draw conclusions in Sec.~\ref{sec:DscCcl}.

\section{Search for New Physics at the precision frontier} \label{sec:Frontier}

High-precision experiments and calculations on simple atomic systems provide a path to test proposed extensions of the SM, complementing particle accelerators, beam dumps and cosmological observations. NP is expected to provide small--and heretofore unobserved--contributions to spectroscopic atomic quantities such as energy levels and $g$ factors. The combined precision of experiment and theory should allow for the setting of competitive bounds on such hypothetical contributions and, hence, on the unknown parameters of SM extensions.\\

This general idea can be implemented directly, by comparing the most precise measurements and calculations of a given quantity, and setting the maximum discrepancy between the two allowed by the error bars to be the maximum possible NP contribution to that quantity~\cite{CoulombHidden,LowEFrontier,FifthForceG}. It can also be implemented indirectly, by considering isotope shifts of spectroscopic quantities~\cite{FifthForce,PossibleForces,RehbehnCoronal}, and scrutinising experimental King plots for deviations from linearity. Such nonlinear King data can be a sign of NP, but it was shown in Refs.~\cite{FifthForceG,FlambaumIso,VladimirKing} that carefully accounting for SM sources of King nonlinearity is important for a reliable interpretation of isotope shift data on few-electron ions. In the present work, the focus is on computing the NP correction to the energy levels of few-electron ions with improved accuracy.

\section{Leading Fifth Force corrections to the energy levels} \label{sec:Leading}

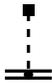
\begin{figure}[bt]
\begin{center}
  \begin{tikzpicture}[very thick,scale=.625]
    \draw (-3.25,-.075) -- (-2.25,-.075);
    \draw (-3.25,.075) -- (-2.25,.075);

    \fill (-2.75,0) circle (.1);

    \draw[dashed] (-2.75,.125)  -- (-2.75,1.25);


    \fill (-2.875,1.5) -- (-2.875,1.25) -- (-2.625,1.25) -- (-2.625,1.5) -- (-2.875,1.5);
    








  \end{tikzpicture}
\end{center}  
  \vspace{-15pt}
  \caption{Feynman diagram corresponding to the leading one-electron contribution to the hypothetical fifth force correction to the energy level of a bound electron. The double line represents the bound electron, and the dashed line terminated by a square denotes the fifth force potential. \label{fig:LeadDiagram}}
\end{figure}

It has been hypothesised~\cite{EightPage} that new massive scalar bosons could carry a new fundamental force, a fifth force (FF), resulting, as far as atomic physics is concerned, in an interaction between nucleons and electrons~\cite{FifthForce,FlambaumIso,ProbingIS}. The spin-independent potential exerted on electrons by this force is of the Yukawa type~\cite{PossibleForces}:
\begin{equation} \label{eq:FFPotential}
  V_{\mathrm{FF}}\left(\mathbf{r}\right)=-\hbar c\,\alpha_{\mathrm{FF}}\,A\,\frac{\mathrm{e}^{-\frac{m_\phi c}{\hbar}\left|\mathbf{r}\right|}}{\left|\mathbf{r}\right|},
\end{equation}
where $m_\phi$ is the mass of the scalar boson that mediates the FF, $\hbar$ and $c$ are Planck's reduced constant and the vacuum velocity of light, and $A$ is the nuclear mass number of the considered ion. This force can be caused by the Higgs portal mechanism~\cite{PossibleForces,HiggsLikeAtomic}, in which case the FF coupling constant reads $\alpha_{\mathrm{FF}}=y_ey_n/4\pi$, with $y_e$ and $y_n$ the coupling of the FF boson to the electrons and the nucleons, respectively. Or, this force can be caused by the gauging of the $B-L$ symmetry in the SM~\cite{PossibleForces}, in which case the FF coupling constant reads $\alpha_{\mathrm{FF}}=g_{B-L}^2/4\pi$, with $g_{B-L}$ the coupling of the new gauge boson to fermions.

 It will be useful to know the momentum-space expression of the Yukawa potential~(\ref{eq:FFPotential}), which reads
\begin{equation} \label{eq:FFPotentialFourier}
  V_{\mathrm{FF}}\left(\mathbf{r}\right)=-\hbar c\,4\pi\,\alpha_{\mathrm{FF}}\,A\,\frac{1}{\mathbf{k}^2+\left(\frac{m_\phi c}{\hbar}\right)^2}.
\end{equation}
The leading FF correction to the energy level $a$ of an ion can be represented by the diagram in Fig.~\ref{fig:LeadDiagram}, and the corresponding expression is simply given by the diagonal matrix element $\bra{a}\hat{V}_{\mathrm{FF}}\ket{a}$ of the FF potential, namely
\begin{equation} \label{eq:ECorr}
  E_{\mathrm{FF}\left(a\right)}=-\alpha_{\mathrm{FF}}\,A\,\hbar c\int_0^{+\infty}\mathrm{d}r\,r\,\mathrm{e}^{-\frac{m_\phi c}{\hbar}r}\left[g_a^2\left(r\right)+f_a^2\left(r\right)\right].
\end{equation}
Here $g_a$ and $f_a$ are the radial wave functions (large and small component, respectively) of the bound electron in state $a$~\cite{Drake}. We give the explicit expression for the $1s$ state in the pointlike nucleus approximation:
\begin{multline} \label{eq:ENP1s}
  E_{\mathrm{FF}\left(1s\right)}=-\alpha_{\mathrm{FF}}\,A\,m_e\,c^2\,\frac{\left(Z\alpha\right)}{\gamma}\left(1+\frac{m_\phi}{2Z\alpha m_e}\right)^{-2\gamma},
\end{multline}
with $m_e$ the electron mass, $\alpha$ the fine-structure constant, and $Z$ the nuclear charge, and $\gamma=\sqrt{\kappa^2-\left(Z\alpha\right)^2}$, with $\kappa$ the relativistic angular quantum number ($\kappa_s = -1$ and $\kappa_{p_{1/2}} = 1$). The expression for the $2s$ and $2p_{1/2}$ states is too heavy to be reproduced here, but can be calculated from Eq.~(\ref{eq:ECorr}). Numerical values for the correction to the $1s$, $2s$ and $2p_{1/2}$ energy levels in single-electron ions are given in Tables~\ref{Tab:Rad_FF_1s}, \ref{Tab:Rad_FF_2s} and \ref{Tab:Rad_FF_2pmin} for various values of the boson mass.
  
\section{Subleading Fifth Force corrections to the energy levels due to QED} \label{sec:Energy}

We now turn to calculating corrections to this FF contribution to the energy levels of few-electron ions. Two families of corrections coming from QED are considered and calculated in detail: the corrections coming from IEI via single photon exchange, and the radiative QED corrections at the one-loop level.

\subsection{Interelectronic interaction corrections} \label{subsec:EnergyXch}

 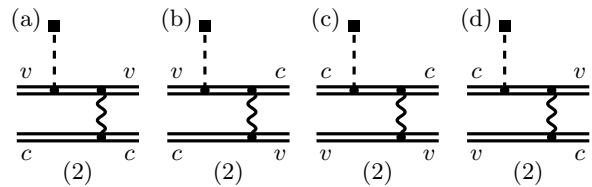
\begin{figure}[b]
  \begin{tikzpicture}[very thick,scale=.625]
    \draw (0.2,1-.075) -- (2.8,1-.075);
    \draw (0.2,1+.075) -- (2.8,1+.075);

    \fill (1,1) circle (.1);
    \fill (2,1) circle (.1);
    
	\draw[dashed] (1,1) -- (1,1+1.25);
        \fill (.875,1.25+1) -- (.875,1.25+1.25) -- (1.125,1.25+1.25) -- (1.125,1.25+1) -- (.875,1.25+1);

    \fill (2,0) circle (.1);
	\draw[decorate,decoration={snake,amplitude=1.5,segment length=6.25}] (2,1) -- (2,0);
    \draw (0.2,-.075) -- (2.8,-.075);
    \draw (0.2,.075) -- (2.8,.075);

    \draw (0.4,1.375) node {$v$};
    \draw (2.6,1.375) node {$v$};
    \draw (0.4,-0.375) node {$c$};
    \draw (2.6,-0.375) node {$c$};

    \draw (0.4,2.5) node {(a)};
    \draw (1.5,-.75) node {(2)};
 \end{tikzpicture}
  \begin{tikzpicture}[very thick,scale=.625]
    \draw (0.2,1-.075) -- (2.8,1-.075);
    \draw (0.2,1+.075) -- (2.8,1+.075);

    \fill (1,1) circle (.1);
    \fill (2,1) circle (.1);
    
	\draw[dashed] (1,1) -- (1,1+1.25);
        \fill (.875,1.25+1) -- (.875,1.25+1.25) -- (1.125,1.25+1.25) -- (1.125,1.25+1) -- (.875,1.25+1);

    \fill (2,0) circle (.1);
	\draw[decorate,decoration={snake,amplitude=1.5,segment length=6.25}] (2,1) -- (2,0);
    \draw (0.2,-.075) -- (2.8,-.075);
    \draw (0.2,.075) -- (2.8,.075);

    \draw (0.4,1.375) node {$v$};
    \draw (2.6,1.375) node {$c$};
    \draw (0.4,-0.375) node {$c$};
    \draw (2.6,-0.375) node {$v$};

    \draw (0.4,2.5) node {(b)};
    \draw (1.5,-.75) node {(2)};
 \end{tikzpicture}
  \begin{tikzpicture}[very thick,scale=.625]
    \draw (0.2,1-.075) -- (2.8,1-.075);
    \draw (0.2,1+.075) -- (2.8,1+.075);

    \fill (1,1) circle (.1);
    \fill (2,1) circle (.1);
    
	\draw[dashed] (1,1) -- (1,1+1.25);
        \fill (.875,1.25+1) -- (.875,1.25+1.25) -- (1.125,1.25+1.25) -- (1.125,1.25+1) -- (.875,1.25+1);

    \fill (2,0) circle (.1);
	\draw[decorate,decoration={snake,amplitude=1.5,segment length=6.25}] (2,1) -- (2,0);
    \draw (0.2,-.075) -- (2.8,-.075);
    \draw (0.2,.075) -- (2.8,.075);

    \draw (0.4,1.375) node {$c$};
    \draw (2.6,1.375) node {$c$};
    \draw (0.4,-0.375) node {$v$};
    \draw (2.6,-0.375) node {$v$};

    \draw (0.4,2.5) node {(c)};
    \draw (1.5,-.75) node {(2)};
 \end{tikzpicture}
  \begin{tikzpicture}[very thick,scale=.625]
    \draw (0.2,1-.075) -- (2.8,1-.075);
    \draw (0.2,1+.075) -- (2.8,1+.075);

    \fill (1,1) circle (.1);
    \fill (2,1) circle (.1);
    
	\draw[dashed] (1,1) -- (1,1+1.25);
        \fill (.875,1.25+1) -- (.875,1.25+1.25) -- (1.125,1.25+1.25) -- (1.125,1.25+1) -- (.875,1.25+1);

    \fill (2,0) circle (.1);
	\draw[decorate,decoration={snake,amplitude=1.5,segment length=6.25}] (2,1) -- (2,0);
    \draw (0.2,-.075) -- (2.8,-.075);
    \draw (0.2,.075) -- (2.8,.075);

    \draw (0.4,1.375) node {$c$};
    \draw (2.6,1.375) node {$v$};
    \draw (0.4,-0.375) node {$v$};
    \draw (2.6,-0.375) node {$c$};

    \draw (0.4,2.5) node {(d)};
    \draw (1.5,-.75) node {(2)};
 \end{tikzpicture}
    \caption{The diagrams corresponding to the one-photon exchange interelectronic-interaction corrections to the fifth force contribution to the energy level of the valence electron. 
Here, $v$ stands for a valence electron, and $c$ for a core electron. All diagrams have an equivalent diagram, as such, their contributions should be counted twice, as indicated by the $2$ between brackets under them.
\label{fig:DiagramIEIEner}}
\end{figure}

The correction to the NP contribution to the energy levels of few-electron ions, due to the interaction between valence and core electrons, can be represented by the sum of the contributions from the diagrams in Fig.~\ref{fig:DiagramIEIEner}, namely:
\begin{widetext}
\begin{subequations} \label{eq:IEI}
\begin{equation}\label{eq:IEIEner}
E_{\mathrm{FF}\left(a\right)}^{\rm IEI} = 
E_{\mathrm{FF}\left(a\right)}^{\rm IEI,a} + E_{\mathrm{FF}\left(a\right)}^{\rm IEI,b} 
+ E_{\mathrm{FF}\left(a\right)}^{\rm IEI,c} + E_{\mathrm{FF}\left(a\right)}^{\rm IEI,d}
\end{equation}
with the contributions of the four diagrams (counted together with those of their respective equivalent diagrams) given by
\begin{align}\label{eq:IEIEnerabcd}
E_{\mathrm{FF}\left(a\right)}^{\rm IEI,a}&= 2 \sum\limits_{n,\epsilon_n \neq \epsilon_v}\frac{\langle v|\hat{V}_{\rm FF}|n\rangle \langle cn|\hat{I}\left(0\right)|cv\rangle}{\epsilon_v-\epsilon_n} \notag \\
E_{\mathrm{FF}\left(a\right)}^{\rm IEI,b} &= -2 \sum\limits_{n,\epsilon_n \neq \epsilon_v}\frac{\langle v|\hat{V}_{\rm FF}|n\rangle \langle cn|\hat{I}\left(\Delta\right)|vc\rangle}{\epsilon_v-\epsilon_n}  - \langle v|\hat{V}_{\rm FF}|v\rangle \langle cv|\hat{I}'\left(\Delta\right)|vc\rangle \notag \\
E_{\mathrm{FF}\left(a\right)}^{\rm IEI,c}&= 2 \sum\limits_{n,\epsilon_n \neq \epsilon_c}\frac{\langle c|\hat{V}_{\rm FF}|n\rangle \langle vn|\hat{I}\left(0\right)|vc\rangle}{\epsilon_c-\epsilon_n} \notag \\
E_{\mathrm{FF}\left(a\right)}^{\rm IEI,d}&= -2 \sum\limits_{n,\epsilon_n \neq \epsilon_c}\frac{\langle c|\hat{V}_{\rm FF}|n\rangle \langle vn|\hat{I}\left(\Delta\right)|cv\rangle}{\epsilon_c-\epsilon_n}  + \langle c|\hat{V}_{\rm FF}|c\rangle \langle cv|\hat{I}'\left(\Delta\right)|vc\rangle.
\end{align}
\end{subequations}
%
Here, $v$ stands for a valence electron in state $\ket{a}$, $c$ for a core electron, $\Delta = \epsilon_v - \epsilon_c$, and $\hat{I}\left(\omega\right)$ is the photon propagator in the Feynman gauge, given in configuration space representation by
\begin{align} \label{eq:IProp}
I\left(\omega,\mathbf{x}_1,\mathbf{x}_2\right) = \alpha \frac{\left(1-\balpha_1\cdot\balpha_2\right)
\exp{\left(\mathrm{i}\,x_{12}\sqrt{\omega^2+\mathrm{i} 0}\right)}}
{4\pi x_{12}},
\end{align}
with the relative distance $x_{12} = \left|\mathbf{x}_1-\mathbf{x}_2\right|$. We also wrote $\hat{I}' = \partial \hat{I}/\partial \omega$, and note that the summation in Eq.~\eqref{eq:IEIEnerabcd} goes over the full spectrum of the Dirac equation, including negative- and positive-energy states. Using the angular decomposition of $I\left(\omega\right)$, as was done for instance in Refs.~\cite{ForVertexF,SelfEnergyWFNat}, and performing the angular integration in Eq.~\eqref{eq:IEIEnerabcd} analytically, we obtain
\begin{align}\label{eq:IEIEnerAng}
 E_{\mathrm{FF}\left(a\right)}^{\rm IEI}  
&= 2 \sqrt{\frac{2j_c+1}{2j_v+1}} R_0\left(cvc\delta_{\rm FF}v, 0\right) 
- 2 \sum\limits_J \frac{\left(-1\right)^{j_v-j_c-J}}{2j_v+1} \left[ {R_J\left(vcc\delta_{\rm FF}v,\Delta\right) + R'_J\left(vccv, \Delta\right)\bra{v}\hat{V}_{\rm FF}\ket{v}} \right] \notag \\
&
 + 2 \sqrt{\frac{2j_c+1}{2j_v+1}} R_0\left(vcv\delta_{\rm FF}c,0\right)
- 2 \sum\limits_J \frac{\left(-1\right)^{j_c-j_v-J}}{2j_v+1} \left[ {R_J\left(cvv\delta_{\rm FF}c,\Delta\right) - R'_J\left(cvvc,\Delta\right)\bra{c}\hat{V}_{\rm FF}\ket{c}} \right].
\end{align}
\end{widetext}
Here, $\delta_{\rm FF}a$ stands for the FF-perturbed state:
\begin{equation} \label{eq:FF_WF}
|\delta_{\rm FF}a\rangle = \sum\limits_{i,\epsilon_i \neq \epsilon_a} 
  \frac{|i\rangle \langle i | \hat{V}_{\rm FF} | a\rangle}{\epsilon_a-\epsilon_i},
\end{equation}
and $R_J\left(abcd,\omega\right)$ and $R'_J\left(abcd,\omega\right)$ are the generalized Slater radial integral given explicitly in Refs.~\cite{ForVertexF,SelfEnergyWFNat} and its derivative w.r.t.~$\omega$, respectively. The radial integrations in Eq.~(\ref{eq:IEIEnerAng}) and the summations over the spectrum, were performed numerically. The calculations are performed in the dual kinetic balance (DKB) approach~\cite{Shabaev2004} based on $B$ splines~\cite{FiniteSpline}. The solutions of the Dirac equation for an arbitrary spherically symmetric potential can be found in a finite size cavity, which allows for the description of both the discrete and continuous spectra with a finite number of electronic states for every given $\kappa$. The wave functions of low-lying bound states, such as those considered here, are very well reproduced, so that summations over the Dirac spectrum can be performed with high accuracy. A homogeneously charged sphere model was used for the charge distribution inside of the nucleus, with the root-mean-square nuclear radii taken from Ref.~\cite{NuclRadiiUpdate}. The final numerical results are given in Sec.~\ref{subsec:ResXch}.

\subsection{Radiative corrections} \label{subsec:EnergyRad}

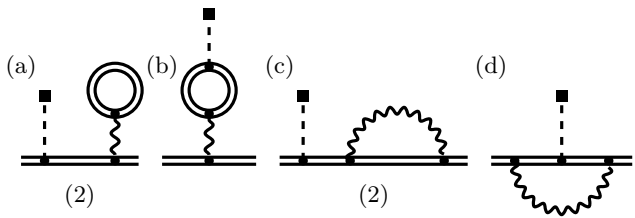
\begin{figure}[tb]
  \begin{tikzpicture}[very thick,scale=.625]
    \draw (-7,-.075) -- (-4.5,-.075);
    \draw (-7,.075) -- (-4.5,.075);

    \fill (-6.5,0) circle (.1);
    \fill (-5,0) circle (.1);
    
        \fill (-7.5+.875,.25+1) -- (-7.5+.875,.25+1.25) -- (-7.5+1.125,.25+1.25) -- (-7.5+1.125,.25+1) -- (-7.5+.875,.25+1);

	\draw[dashed] (-6.5,0) -- (-6.5,1.25);

    \draw[decorate,decoration={snake,amplitude=1.5,segment length=6.25}] (-5,.125)  -- (-5,1);

    \fill (-5,1) circle (.1);

    \draw (-5,1+.5) circle (1.15/2);
    \draw (-5,1+.5) circle (.85/2);




    \draw (-7,2) node {(a)};
    \draw (-5.75,-.75) node {(2)};
    \draw (-4,-.075) -- (-2,-.075);
    \draw (-4,.075) -- (-2,.075);

    \fill (-3+0,0) circle (.1);
    \fill (-3,2) circle (.1);
    
    \fill (-4+.875,.25+3) -- (-4+.875,.25+2.75) -- (-4+1.125,.25+2.75) -- (-4+1.125,.25+3) -- (-4+.875,.25+3);

    \draw[dashed] (-3,2) -- (-3,3);

    \draw[decorate,decoration={snake,amplitude=1.5,segment length=6.25}] (-3+0,.125)  -- (-3+0,1);

    \fill (-3,1) circle (.1);

    \draw (-3,1+.5) circle (1.15/2);
    \draw (-3,1+.5) circle (.85/2);




    \draw (-4,2) node {(b)};
    \draw (-1.5,-.075) -- (2.5,-.075);
    \draw (-1.5,.075) -- (2.5,.075);

    \fill (-1,0) circle (.1);
    
    \fill (-2+.875,.25+1) -- (-2+.875,.25+1.25) -- (-2+1.125,.25+1.25) -- (-2+1.125,.25+1) -- (-2+.875,.25+1);

    \draw[dashed] (-1,0) -- (-1,1.25);







    \draw[decorate,decoration={complete sines,amplitude=2.75,segment length=5.}] (0,0) arc (180:0:2);

    \fill (0,0) circle (.1);
    \fill (2,0) circle (.1); 

    \draw (-1.5,2) node {(c)};
    \draw (.5,-.75) node {(2)};
    \draw (3.5-.5,-.075) -- (3.5+2.5,-.075);
    \draw (3.5-.5,.075) -- (3.5+2.5,.075);

    \fill (3.5+1,0) circle (.1);
    
     \fill (3.5+.875,.25+1) -- (3.5+.875,.25+1.25) -- (3.5+1.125,.25+1.25) -- (3.5+1.125,.25+1) -- (3.5+.875,.25+1);

    \draw[dashed] (4.5,0) -- (4.5,1.25);







    \draw[decorate,decoration={complete sines,amplitude=2.75,segment length=5.}] (3.5+0,0) arc (180:360:2);

    \fill (3.5+0,0) circle (.1);
    \fill (3.5+2,0) circle (.1); 

    \draw (3,2) node {(d)};
 \end{tikzpicture}
  \vspace{-25pt}
    \caption{The diagrams corresponding to the one-loop radiative corrections to the fifth force contribution to the energy level of the valence electron. The diagrams are referred to as (a) electric loop and (b) bosonic loop vacuum polarization contributions, and as the (c) wave function-type and (d) vertex-type self-energy contributions. Diagrams (a) and (c) each have an equivalent diagram, as such, their contributions should be counted twice, as indicated by the $2$ between brackets under them. \label{fig:DiagramRadEner}}
\end{figure}

The one-loop radiative corrections to the hypothetical FF contribution to the energy levels of bound valence electrons are given by the contributions of the diagrams represented in Fig.~\ref{fig:DiagramRadEner}. Diagram~\ref{fig:DiagramRadEner}(a) corresponds to the electric loop correction, \ref{fig:DiagramRadEner}(b) to the bosonic loop correction (named in analogy with the familiar electric loop and magnetic loop diagrams~\cite{DelbrueckGArticle}). They are both vacuum polarization (VP) corrections. In this work, we treat fermionic VP loops in the free-loop approximation. It is expected~\cite{VladZVP,CzarneckiLetter,DelbrueckGArticle} that further binding corrections to this approximation only bring about smaller contributions, which are hence of little relevance to the search for NP. Diagrams \ref{fig:DiagramRadEner}(c) and \ref{fig:DiagramRadEner}(d) correspond to the self-energy (SE) correction.

\subsubsection{Vacuum polarization corrections} \label{subsubsec:EnergyRadVP}

The correction to the energy corresponding to the electric loop diagram~\ref{fig:DiagramRadEner}(a) is given by
\begin{equation} \label{eq:EDiaga}
E_{\mathrm{FF}\left(a\right)}^{\mathrm{a}}=2\sum_{i,\epsilon_i\neq\epsilon_a}\frac{\bra{a}\hat{V}_{\mathrm{FF}}\ket{i}\bra{i}\hat{V}_{\mathrm{VP}}\ket{a}}{\epsilon_a-\epsilon_i},
\end{equation}
where $\hat{V}_{\mathrm{VP}}$ is the vacuum polarization (electric loop) operator. It is then convenient to numerically compute this sum by using either the FF-corrected wave functions, or the VP-corrected wave functions. Both approaches evidently yield the same result. In the latter approach, the energy correction can be rewritten as
\begin{multline} \label{eq:EDiagaPart}
E_{\mathrm{FF}\left(a\right)}^{\mathrm{a}}=-2\alpha_{\mathrm{FF}}\,A\,\hbar c\int_0^{+\infty}\mathrm{d}r\,r\,\mathrm{e}^{-\frac{m_\phi c}{\hbar}r}\\
\times\left[g_a\left(r\right)\left(\delta_{\mathrm{VP}}g_a\left(r\right)\right)+f_a\left(r\right)\left(\delta_{\mathrm{VP}}f_a\left(r\right)\right)\right],
\end{multline}
where $\delta_{\mathrm{VP}}g_a$ and $\delta_{\mathrm{VP}}f_a$ are the corrections to the radial wave functions due to the VP loop (defined in the same way as in Eq.~(\ref{eq:FF_WF}), but with the VP potential). In the free-loop approximation used here, the correction is generated by the Uehling potential~\cite{Uehling,UehlingAnal}
\begin{widetext}
\begin{equation} \label{eq:UehlingPot}
V_{\mathrm{Ue}}\left(r\right)=-\frac{Z\alpha}{r}\,\hbar c\left(\frac{2\alpha}{3\pi}\right)\int_1^{+\infty}\frac{\mathrm{d}u}{u^2}\,\sqrt{u^2-1}\left(1+\frac{1}{2u^2}\right)\mathrm{e}^{-2u\frac{m_e c}{\hbar}r},
\end{equation}
and the wave function corrections can be computed numerically using known methods.

The contribution from the bosonic loop diagram~\ref{fig:DiagramRadEner}(b) is simply given by the matrix element
\begin{equation} \label{eq:EDiagb}
E_{\mathrm{FF}\left(b\right)}^{\mathrm{b}}=\bra{a}\hat{V}_{\mathrm{BL}}\ket{a},
\end{equation}
where the bosonic loop potential $\hat{V}_{\mathrm{BL}}$ comes from the insertion of a VP loop into the propagator of the FF boson in the tree-level diagram of Fig.~\ref{fig:LeadDiagram}. We note that the propagator between the VP loop and the bound electron in diagram~\ref{fig:DiagramRadEner}(b) is that of a photon. If it were that of a FF boson, the corresponding correction would be quadratic in the FF coupling constant $\alpha_{\mathrm{FF}}$ and hence presumably many orders of magnitude smaller than all contributions considered in this work. The bosonic loop potential can be derived following the same method used for deriving the Uehling potential (see details in App.~\ref{appdx:BLoop}). It takes a similar, but somewhat more involved form, namely
\begin{equation} \label{eq:BosonLoopPot}
V_{\mathrm{BL}}\left(r\right)=-\frac{A\alpha}{r}\,\hbar c\left(\frac{2\alpha_{\mathrm{FF}}}{3\pi}\right)\int_1^{+\infty}\frac{\mathrm{d}u}{u^2}\,\sqrt{u^2-1}\left(1+\frac{1}{2u^2}\right)\left[\frac{\mathrm{e}^{-2u\frac{m_e c}{\hbar}r}-\left(\frac{m_\phi}{2um_e}\right)^2\mathrm{e}^{-\frac{m_\phi c}{\hbar}r}}{1-\left(\frac{m_\phi}{2um_e}\right)^2}\right].
\end{equation}
\end{widetext}
Although it is not obvious at first glance, the integrand is regular at $u=m_\phi/\left(2m_e\right)$. Expressions of the Uehling potential (\ref{eq:UehlingPot}) convenient for numerical implementation were given by Klarsfeld in Ref.~\cite{Klarsfeld1977} in terms of modified Bessel functions. We used this approach to calculate the Uehling potential (\ref{eq:UehlingPot}) and, thence, the electric loop correction (\ref{eq:EDiaga}) to the energy levels. On the other hand, this approach cannot be generalized to the bosonic loop potential (\ref{eq:BosonLoopPot}). Indeed, the corresponding expression would involve an infinite sum of terms to be re-expressed as modified Bessel functions, and in Ref.~\cite{Klarsfeld1977}, the lower-index Bessel terms are expressed in function of the highest-index relevant term, which does not exist for an infinite sum, and cannot be determined in advance for the truncated sums used for numerical implementation. As a result, we numerically compute the bosonic loop potential on pre-determined radial grids.

\subsubsection{Self-energy corrections} \label{subsubsec:EnergyRadSE}

In the same way as what was done in Ref.~\cite{YeroGFactor} for the SE corrections to the $g$ factor, the contribution of diagram~\ref{fig:DiagramRadEner}(c) to the energy level $a$ can be decomposed in two contributions, referred to as reducible and irreducible, respectively. The propagator of the bound electron that is found between the FF vertex and the self-energy loop can be written in spectral form:
\begin{equation} \label{eq:GreenFct}
\hat{G}\left(\epsilon\right)=\sum_k\frac{\ket{k}\bra{k}}{\left(\epsilon-\epsilon_k\right)},
\end{equation}
that is, as a sum over the bound and continuous parts of the Dirac-Coulomb spectrum. The term for which $k=a$ (the intermediate state coincides with the reference state) generates the so-called reducible contribution, which is given by
\begin{equation} \label{eq:EDiagcRED}
E_{\mathrm{FF}\left(a\right)}^{\mathrm{c}\left(\mathrm{red}\right)}=E_{\mathrm{FF}\left(a\right)}\bra{a}\gamma^0\left.\frac{\partial\hat{\Sigma}}{\partial\epsilon}\right|_{\epsilon=\epsilon_a}\ket{a},
\end{equation}
with $E_{\mathrm{FF}\left(a\right)}$ the leading one-electron FF correction (\ref{eq:ECorr}) to the energy of level $a$ and $\hat{\Sigma}\left(\epsilon\right)$ the self-energy operator, studied in detail in Ref.~\cite{ForVertexF}.

All other terms in the bound electron propagator, for which $k\neq a$, are added together to generate the so-called irreducible contribution. This term can be computed either as
\begin{subequations} \label{eq:EDiagcIRR}
\begin{equation} \label{eq:EDiagcIRR_FF}
E_{\mathrm{FF}\left(a\right)}^{\mathrm{c}\left(\mathrm{irr}\right)}=\bra{\delta_{\mathrm{FF}}a}\gamma^0\hat{\widetilde{\Sigma}}\left(\epsilon_a\right)\ket{a}+\bra{a}\gamma^0\hat{\widetilde{\Sigma}}\left(\epsilon_a\right)\ket{\delta_{\mathrm{FF}}a}
\end{equation}
where $\ket{\delta_{\mathrm{FF}}a}$ is the correction (\ref{eq:FF_WF}) to the bound state $a$ from the FF potential (\ref{eq:FFPotential}), or as
\begin{equation} \label{eq:EDiagcIRR_SE}
E_{\mathrm{FF}\left(a\right)}^{\mathrm{c}\left(\mathrm{irr}\right)}=\bra{\delta_{\mathrm{SE}}a}\hat{V}_{\mathrm{FF}}\ket{a}+\bra{a}\hat{V}_{\mathrm{FF}}\ket{\delta_{\mathrm{SE}}a}
\end{equation}
\end{subequations}
where $\ket{\delta_{\mathrm{SE}}a}$ is the SE correction to the bound state $a$. A method has been developed for the challenging numerical computation of the SE-corrected wave functions, and described in detail in Ref.~\cite{SelfEnergyWFNat}. We use the codes and results developed in that work, for our present calculations. The matrix elements (\ref{eq:EDiagcIRR_SE}) can be obtained through an expression identical to Eq.~(\ref{eq:EDiagaPart}), with the VP-corrected wave functions replaced with the SE-corrected ones.

The contribution of diagram~\ref{fig:DiagramRadEner}(d) (the so-called vertex diagram) is decomposed into a UV-divergent zero-potential term, where the propagator of the electron under the self-energy loop is taken to be that of a free electron, and a finite many-potential term, according to the standard method~\cite{BeierOneLoop,YeroGFactor,BastianPhD}. The UV divergence is cancelled by a divergence in the reducible contribution (\ref{eq:EDiagcRED}), and the renormalized zero-potential term reads
\begin{widetext}
\begin{equation} \label{eq:EDiagdZER}
E_{\mathrm{FF}\left(a\right)}^{\mathrm{d}\left(0\right)}=\int\frac{\mathrm{d}\mathbf{p}}{\left(2\pi\right)^3}\int\frac{\mathrm{d}\mathbf{p}'}{\left(2\pi\right)^3}\,\bar{\psi}_a\left(\mathbf{p}\right)\Gamma_R^0\left(p,p'\right)V_{\mathrm{FF}}\left(\mathbf{p}-\mathbf{p}'\right)\psi_a\left(\mathbf{p}'\right)
\end{equation}
where $\Gamma_R$ is the renormalized vertex function, studied in detail in Ref.~\cite{ForVertexF}, and the four-momenta read $p=\left(\epsilon_a/c,\mathbf{p}\right)$, $p'=\left(\epsilon_a/c,\mathbf{p}'\right)$. The many-potential term is expressed as
\begin{equation} \label{eq:EDiagdMNY}
E_{\mathrm{FF}\left(a\right)}^{\mathrm{d}\left(1+\right)}=\frac{\mathrm{i}}{2\pi}\int_{-\infty}^{+\infty}\mathrm{d}\omega\sum_{n_1n_2}\frac{\bra{a n_2}\hat{I}\left(\omega\right)\ket{n_1a}\bra{n_1}\hat{V}_{\mathrm{FF}}\ket{n_2}}{\left(\epsilon_a-\omega-\epsilon_{n_1}\left(1-\mathrm{i}\eta\right)\right)\left(\epsilon_a-\omega-\epsilon_{n_2}\left(1-\mathrm{i}\eta\right)\right)}-\text{zero pot.}
\end{equation}
\end{widetext}
Here, the subtraction of the zero-potential term corresponds to subtracting a term structurally identical to the explicitly written one, but with the double sum over bound states replaced with a double sum over the spectrum of the free Dirac equation. The operator $\hat{I}$ was defined in Eq.~(\ref{eq:IProp}), and the calculation of its matrix elements has been carried out following the methods of Refs.~\cite{ForVertexF,YeroGFactor}.

\section{Numerical results} \label{sec:Res}

\begin{figure*}[bt]
\subfloat[$Z=14$ (Si nucleus)\label{fig:Silicon}]{%
  \includegraphics[width=.5\textwidth]{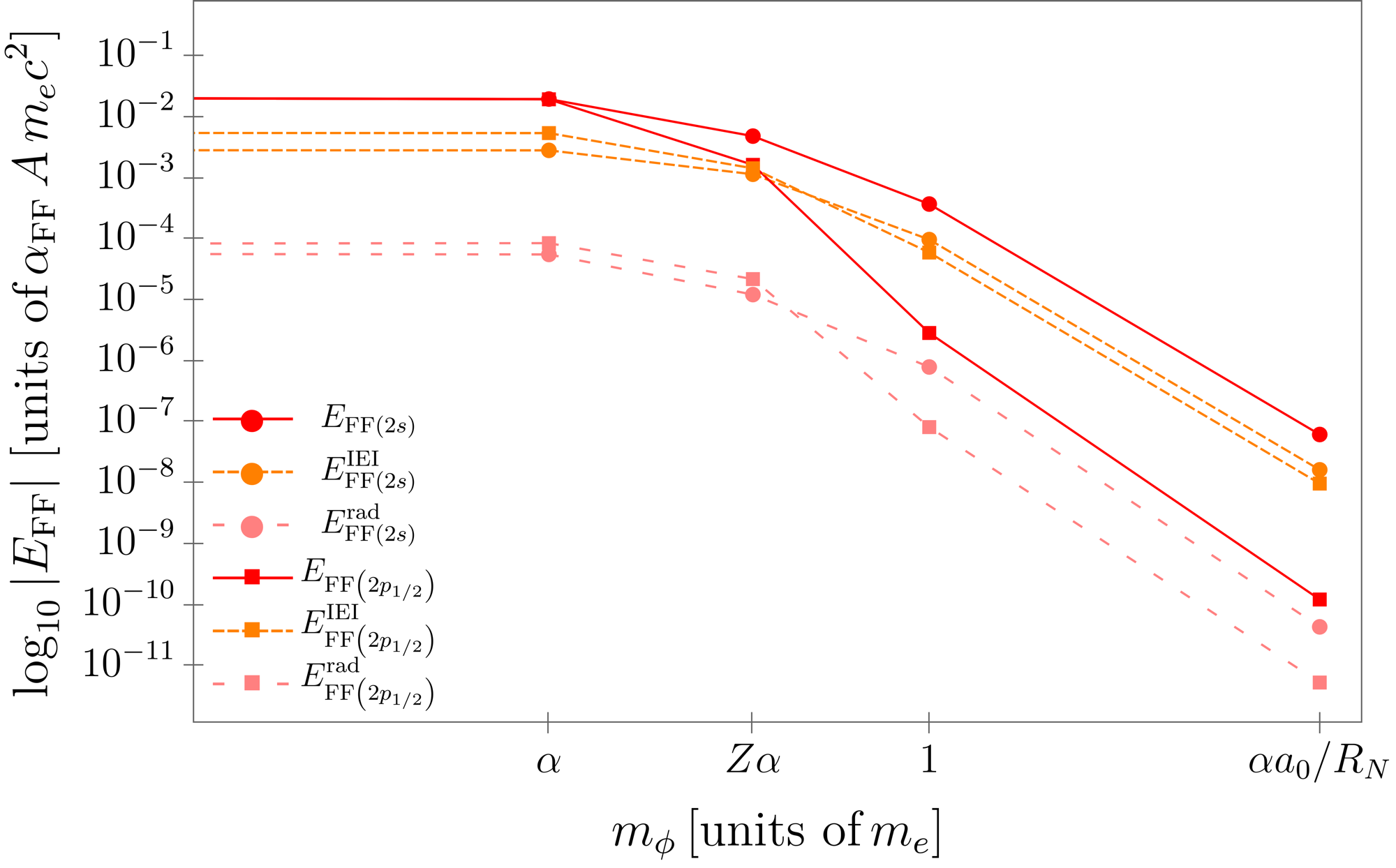}%
}
\hfill
\subfloat[$Z=50$ (Sn nucleus)\label{fig:Tin}]{%
  \includegraphics[width=.5\textwidth]{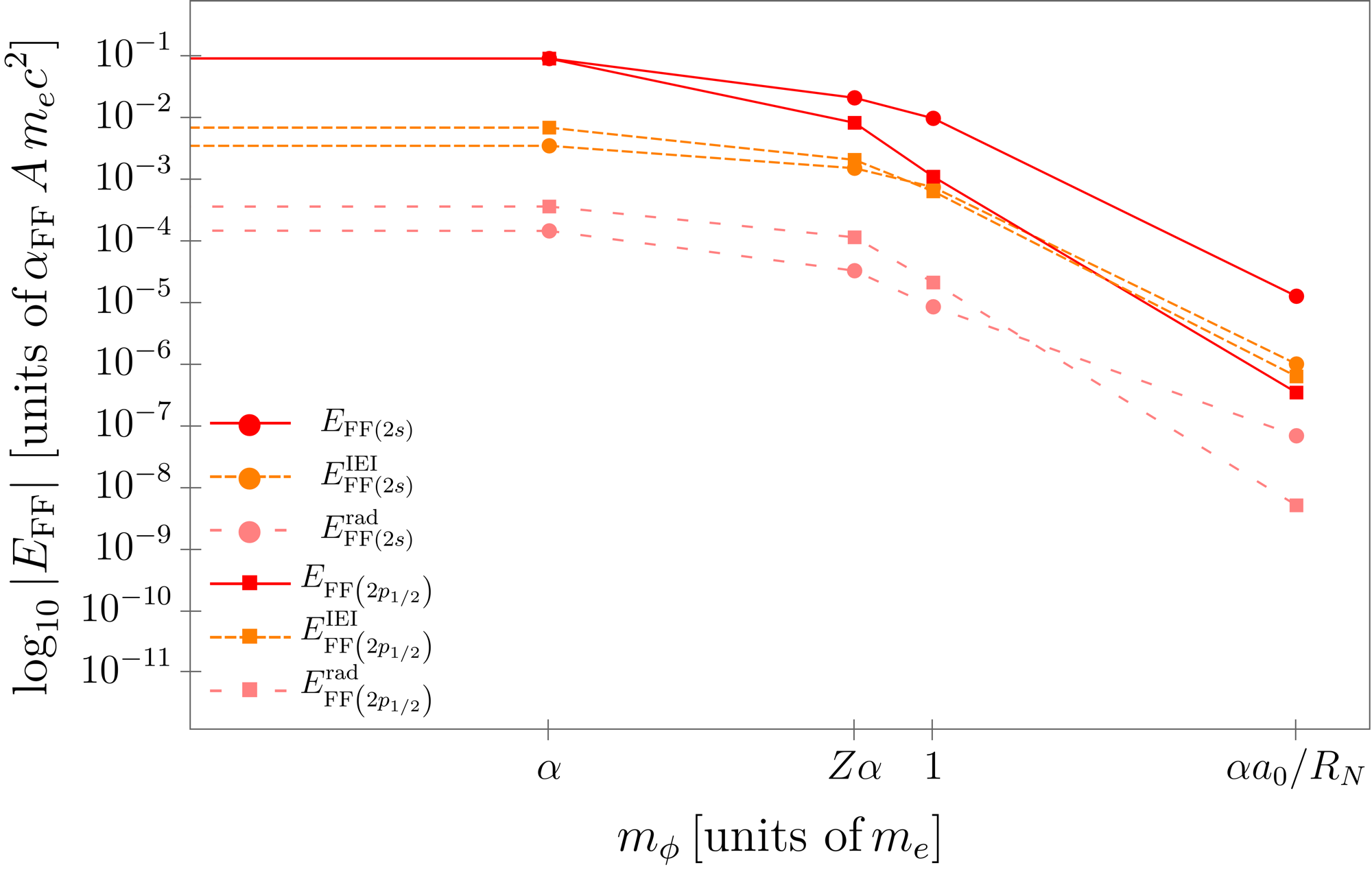}%
}
\caption{Contributions of the hypothetical fifth force to the energy levels of few-electron ions, as a function of the new boson mass $m_\phi$. The leading one-electron contribution is represented in solid red, the contribution from one-photon interelectronic interaction is represented in densely dashed orange, and the radiative contribution at the one-loop level is represented in loosely dashed pink. The values for Li-like ions are represented by circles and those for B-like ions are represented by squares. Both axes are in logarithmic scale. \label{fig:SiSn}}
\end{figure*}

To analyze the importance of the calculated effects for different values of the hypothetical boson mass $m_\phi$ and of the nuclear charge $Z$, we calculated the respective contributions for a broad range of combinations of these parameters. We also give a graphical representation of these results for Li-like and B-like Si and Sn in Fig.~\ref{fig:SiSn}.

\subsection{Interelectronic interaction corrections} \label{subsec:ResXch}

In Table~\ref{Tab:IEI_FF_2s}, we present the results for the IEI correction to the FF contribution to the energy of the $2s$ state of a Li-like ion, giving the individual contributions a, b, c, d, corresponding to the diagrams in Fig.~\ref{fig:DiagramIEIEner}, and the total contribution. In Table~\ref{Tab:IEI_FF_2p}, we present the results for the IEI correction to the FF contribution to the energy of the $2p_{1/2}$ state of a B-like ion. In this Table, we present the individual contributions due to the interactions of the valence electron with the specific core electron shells, as well as the total contribution. All results are given in units of $\alpha_{\mathrm{FF}}\,A\,m_e c^2$.

\begin{table*}[t]
\begin{center}
\begin{tabular}{c@{\hskip 0.1125in}c@{\hskip 0.1125in}c|@{\hskip 0.1125in}c|@{\hskip 0.1125in}l@{\hskip 0.1125in}l@{\hskip 0.1125in}l@{\hskip 0.1125in}l|@{\hskip 0.1125in}l }
\hline \hline
\rule[-3mm]{0mm}{8mm}
$Z$ & $A$ & $R_N$ ($\mathrm{fm}$) & $m_\phi$ & {$E_{\mathrm{FF}(2s)}^{\rm IEI,a}$}\hskip 0.1125in & {$E_{\mathrm{FF}(2s)}^{\rm IEI,b}$}\hskip 0.1125in & {$E_{\mathrm{FF}(2s)}^{\rm IEI,c}$}\hskip 0.1125in & {$E_{\mathrm{FF}(2s)}^{\rm IEI,d}$}\hskip 0.1125in & {$E_{\mathrm{FF}(2s)}^{\rm IEI}$}\hskip 0.1125in \\ 
\hline 
6 & 12 & 2.4702 & $0$ & $2.649\left[-3\right]$ & $1.079\left[-4\right]$ & $4.219\left[-4\right]$ & 	 $-2.680\left[-4\right]$ & $2.911\left[-3\right]$ \\
& & & $\alpha m_e$ & $2.234\left[-3\right]$ & $1.465\left[-4\right]$ & $4.135\left[-4\right]$ & $-2.629\left[-4\right]$ & $2.531\left[-3\right]$\\
& & & $Z\alpha m_e$ & $8.480\left[-4\right]$ & $1.769\left[-4\right]$ & $2.998\left[-4\right]$ & $-1.925\left[-4\right]$ & $1.132\left[-3\right]$\\
& & & $m_e$ &  $1.740\left[-5\right]$ & $2.257\left[-6\right]$ & $5.983\left[-6\right]$ & $-3.884\left[-6\right]$ & $2.176\left[-5\right]$\\
\rule[-3mm]{0mm}{4mm}
& & & $\hbar/R_N c$ & $1.423\left[-9\right]$ & $1.819\left[-10\right]$ & $4.882\left[-10\right]$ & $-3.169\left[-10\right]$ & $1.776\left[-9\right]$\\

14 & 28 & 3.1224 & $0$ & $2.680\left[-3\right]$ & $1.131\left[-4\right]$ & $4.284\left[-4\right]$ & 	 $-2.725\left[-4\right]$ & $2.949\left[-3\right]$ \\
& & & $\alpha m_e$ & $2.567\left[-3\right]$ & $1.238\left[-4\right]$ & $4.267\left[-4\right]$ & $-2.713\left[-4\right]$ & $2.846\left[-3\right]$\\
& & & $Z\alpha m_e$ & $8.651\left[-4\right]$ & $1.804\left[-4\right]$ & $3.052\left[-4\right]$ & $-1.959\left[-4\right]$ & $1.154\left[-3\right]$\\
& & & $m_e$ &  $7.795\left[-5\right]$ & $1.066\left[-5\right]$ & $2.694\left[-5\right]$ & $-1.743\left[-5\right]$ & $9.812\left[-5\right]$\\
\rule[-3mm]{0mm}{4mm}
& & & $\hbar/R_N c$ & $1.321\left[-8\right]$ & $1.692\left[-9\right]$ & $4.521\left[-9\right]$ & $-2.925\left[-9\right]$ & $1.650\left[-8\right]$\\

20 & 40 & 3.4776 & $0$ & $2.720\left[-3\right]$ & $1.197\left[-4\right]$ & $4.370\left[-4\right]$ & 	 $-2.785\left[-4\right]$ & $2.998\left[-3\right]$ \\
& & & $\alpha m_e$ & $2.659\left[-3\right]$ & $1.253\left[-4\right]$ & $4.361\left[-4\right]$ & $-2.778\left[-4\right]$ & $2.943\left[-3\right]$\\
& & & $Z\alpha m_e$ & $8.874\left[-4\right]$ & $1.849\left[-4\right]$ & $3.123\left[-4\right]$ & $-2.002\left[-4\right]$ & $1.184\left[-3\right]$\\
& & & $m_e$ &  $1.398\left[-4\right]$ & $2.016\left[-5\right]$ & $4.861\left[-5\right]$ & $-3.136\left[-5\right]$ & $1.772\left[-4\right]$\\
\rule[-3mm]{0mm}{4mm}
& & & $\hbar/R_N c$ & $3.612\left[-8\right]$ & $4.639\left[-9\right]$ & $1.232\left[-8\right]$ & $-7.944\left[-9\right]$ & $4.514\left[-8\right]$\\
 
50 & 120 & 4.6519 & $0$ & $3.187\left[-3\right]$ & $1.937\left[-4\right]$ & $5.415\left[-4\right]$ & 	 $-3.496\left[-4\right]$ & $3.572\left[-3\right]$ \\
& & & $\alpha m_e$ & $3.175\left[-3\right]$ & $1.943\left[-4\right]$ & $5.412\left[-4\right]$ & $-3.491\left[-4\right]$ & $3.562\left[-3\right]$\\
& & & $Z\alpha m_e$ & $1.160\left[-3\right]$ & $2.369\left[-4\right]$ & $3.991\left[-4\right]$ & $-2.537\left[-4\right]$ &    $1.542\left[-3\right]$\\
& & & $m_e$ &  $5.779\left[-4\right]$ & $1.059\left[-4\right]$ & $2.068\left[-4\right]$ & $-1.305\left[-4\right]$ & $7.601\left[-4\right]$\\
\rule[-3mm]{0mm}{4mm}
& & & $\hbar/R_N c$ & $8.181\left[-7\right]$ & $1.050\left[-7\right]$ & $2.716\left[-7\right]$ & $-1.693\left[-7\right]$ & $1.025\left[-6\right]$\\

92 & 238 	& 5.8337 & $0$ & $5.464\left[-3\right]$ & $4.930\left[-4\right]$ & $1.136\left[-3\right]$ & 	 $-7.379\left[-4\right]$ & $6.355\left[-3\right]$ \\
& & & $\alpha m_e$ & $5.460\left[-3\right]$ & $4.924\left[-4\right]$ & $1.135\left[-3\right]$ & $-7.370\left[-4\right]$ & $6.351\left[-3\right]$\\
& & & $Z\alpha m_e$ & $2.695\left[-3\right]$ & $4.763\left[-4\right]$ & $9.180\left[-4\right]$ & $-5.717\left[-4\right]$ & $3.518\left[-3\right]$\\
& & & $m_e$ &  $2.222\left[-3\right]$ & $4.022\left[-4\right]$ & $7.957\left[-4\right]$ & $-4.916\left[-4\right]$ & $2.928\left[-3\right]$\\
\rule[-3mm]{0mm}{4mm}
& & & $\hbar/R_N c$ & $2.704\left[-5\right]$ & $2.901\left[-6\right]$ & $8.822\left[-6\right]$ & $-5.268\left[-6\right]$ & $3.349\left[-5\right]$\\
\hline \hline
\end{tabular}
\caption{Interelectronic interaction corrections to the hypothetical fifth-force contributions to the $2s$ energy level, for various ions and various new boson masses $m_\phi$. The contributions from the diagrams on Fig.~\ref{fig:DiagramIEIEner} are listed individually, and their sum is given in the last column. All corrections are given in units of $\alpha_{\mathrm{FF}}\,A\,m_e c^2$, the product of the nuclear mass number with the New Physics coupling constant and the electron rest energy. Powers of $10$ are given between square brackets.} 
\label{Tab:IEI_FF_2s}
\end{center}
\end{table*}

\begin{table*}[t]
\begin{center}
\begin{tabular}{c c c | c | l l l}
\hline \hline
\rule[-3mm]{0mm}{8mm}
$Z$ & $A$ & $R_N$ ($\mathrm{fm}$) & $m_\phi$ &  \multicolumn{3}{c}{$E_{\mathrm{FF}(2p_{1/2})}^{\rm IEI}$} \\
\multicolumn{3}{c|}{Core shell} &  & $1s$ & $2s$ & total  \\
\hline 
6 & 12 & 2.4702 & $0$ & $3.431\left[-3\right]$ & $2.157\left[-3\right]$ & $5.589\left[-3\right]$ \\
& & & $\alpha m_e$ & $2.957\left[-3\right]$ & $1.816\left[-3\right]$ & $4.774\left[-3\right]$ \\
& & & $Z\alpha m_e$ & $8.457\left[-4\right]$ & $5.630\left[-4\right]$ & $1.408\left[-3\right]$ \\
& & & $m_e$ 	& $5.671\left[-6\right]$ & $7.318\left[-6\right]$ & $1.299\left[-5\right]$ \\
\rule[-3mm]{0mm}{4mm}
& & & $\hbar/R_N c$ & $4.586\left[-10\right]$ & $5.960\left[-10\right]$ & $1.054\left[-9\right]$ \\

14 & 28 & 3.1224 & $0$ & $3.490\left[-3\right]$ & $2.181\left[-3\right]$ & $5.671\left[-3\right]$ \\
& & & $\alpha m_e$ & $3.368\left[-3\right]$ & $2.090\left[-3\right]$ & $5.459\left[-3\right]$ \\
& & & $Z\alpha m_e$ & $8.711\left[-4\right]$ & $5.747\left[-4\right]$ & $1.445\left[-3\right]$ \\
& & & $m_e$ 	& $2.685\left[-5\right]$ & $3.328\left[-5\right]$ & $6.013\left[-5\right]$ \\
\rule[-3mm]{0mm}{4mm}
& & & $\hbar/R_N c$ & $4.295\left[-9\right]$ & $5.527\left[-9\right]$ & $9.823\left[-9\right]$ \\

20 & 40 & 3.4776 & $0$ & $3.566\left[-3\right]$ & $2.212\left[-3\right]$ & $5.779\left[-3\right]$ \\
& & & $\alpha m_e$ & $3.501\left[-3\right]$ & $2.164\left[-3\right]$ & $5.665\left[-3\right]$ \\
& & & $Z\alpha m_e$ & $9.044\left[-4\right]$ & $5.901\left[-4\right]$ & $1.494\left[-3\right]$ \\
& & & $m_e$ 	& $5.149\left[-5\right]$ & $6.078\left[-5\right]$ & $1.122\left[-4\right]$ \\
\rule[-3mm]{0mm}{4mm}
& & & $\hbar/R_N c$ & $1.188\left[-8\right]$ & $1.509\left[-8\right]$ & $2.697\left[-8\right]$ \\

50 & 120 & 4.6519 & $0$ & $4.469\left[-3\right]$ & $2.582\left[-3\right]$ & $7.051\left[-3\right]$ \\
& & & $\alpha m_e$ & $4.456\left[-3\right]$ & $2.573\left[-3\right]$ & $7.029\left[-3\right]$ \\
& & & $Z\alpha m_e$ & $1.328\left[-3\right]$ & $7.810\left[-4\right]$ & $2.109\left[-3\right]$ \\
& & & $m_e$ 	& $3.588\left[-4\right]$ & $2.901\left[-4\right]$ & $6.489\left[-4\right]$ \\
\rule[-3mm]{0mm}{4mm}
& & & $\hbar/R_N c$ & $3.044\left[-7\right]$ & $3.399\left[-7\right]$ & $6.443\left[-7\right]$ \\

92 & 238 	& 5.8337 & $0$ & $9.206\left[-3\right]$ & $4.474\left[-3\right]$ & $1.368\left[-2\right]$ \\
& & & $\alpha m_e$ & $9.201\left[-3\right]$ & $4.471\left[-3\right]$ & $1.367\left[-2\right]$ \\
& & & $Z\alpha m_e$ & $4.143\left[-3\right]$ & $1.940\left[-3\right]$ & $6.083\left[-3\right]$ \\
& & & $m_e$ 	& $2.953\left[-3\right]$ & $1.458\left[-3\right]$ & $4.411\left[-3\right]$ \\
\rule[-3mm]{0mm}{4mm}
& & & $\hbar/R_N c$ & $1.557\left[-5\right]$ & $1.195\left[-5\right]$ & $2.753\left[-5\right]$ \\
\hline \hline
\end{tabular}
\caption{Interelectronic interaction corrections to the hypothetical fifth-force contributions to the $2p_{1/2}$ energy level, for various ions and various new boson masses $m_\phi$. The contributions from interaction with different core shells are listed separately, and their sum is given in the last column.
All corrections are given in units of $\alpha_{\mathrm{FF}}\,A\,m_e c^2$, the product of the nuclear mass number with the New Physics coupling constant and the electron rest energy. Powers of $10$ are given between square brackets.}
\label{Tab:IEI_FF_2p}
\end{center}
\end{table*}

\subsection{Radiative corrections} \label{subsec:ResRad}

In Tables~\ref{Tab:Rad_FF_1s}, \ref{Tab:Rad_FF_2s} and \ref{Tab:Rad_FF_2pmin}, we respectively present the results for the radiative corrections to the FF contribution to the energy of the $1s$ state of a H-like ion, the $2s$ state of a Li-like ion, and the $2p_{1/2}$ state of a B-like ion, giving the individual contributions a, b, c, d, corresponding to the diagrams in Fig.~\ref{fig:DiagramRadEner}, and the total contribution. All results are given in units of $\alpha_{\mathrm{FF}}\,A\,m_e c^2$.

\begin{table*}[t]
\begin{center}
\begin{tabular}{c@{\hskip 0.075in}c@{\hskip 0.075in}c|@{\hskip 0.05in}c|@{\hskip 0.1125in}l|@{\hskip 0.1125in}l@{\hskip 0.1125in}l@{\hskip 0.1125in}l@{\hskip 0.1125in}l|@{\hskip 0.05in}l}
\hline \hline
\rule[-3mm]{0mm}{8mm}
$Z$ & $A$ & $R_N$ ($\mathrm{fm}$) & $m_\phi$ & \multicolumn{1}{c|}{$E_{\mathrm{FF}\left(1s\right)}$}\hskip 0.1125in & \multicolumn{1}{c}{$E_{\mathrm{FF}\left(1s\right)}^{\mathrm{a}}$}\hskip 0.1125in & \multicolumn{1}{c}{$E_{\mathrm{FF}\left(1s\right)}^{\mathrm{b}}$}\hskip 0.1125in & \multicolumn{1}{c}{$E_{\mathrm{FF}\left(1s\right)}^{\mathrm{c}\left(\mathrm{irr}\right)}$}\hskip 0.1125in & \multicolumn{1}{c|}{$E_{\mathrm{FF}\left(1s\right)}^{\mathrm{c}\left(\mathrm{red}\right)+\mathrm{d}}$}\hskip 0.05in & \multicolumn{1}{c}{$E_{\mathrm{FF}\left(1s\right)}^{\mathrm{rad}}$}\\
\hline 

6 & 12 & 2.4702 & $0$ & $-4.383\left[-2\right]$ & $-1.473\left[-7\right]$ & $-4.972\left[-8\right]$ & $-3.01(2)\left[-6\right]$ & $-3.1(2)\left[-7\right]$ & $-3.96(3)\left[-6\right]$\\ 
& & & $\alpha m_e$ & $-3.735\left[-2\right]$ & $-1.456\left[-7\right]$ & $-4.942\left[-8\right]$ & $-2.97(2)\left[-6\right]$ & $\hphantom{-}1.04(2)\left[-6\right]$ & $-2.13(3)\left[-6\right]$\\
& & & $Z\alpha m_e$ & $-1.949\left[-2\right]$ & $-1.192\left[-7\right]$ & $-4.395\left[-8\right]$ & $-2.39(2)\left[-6\right]$ & $\hphantom{-}8.9(2)\left[-7\right]$ & $-1.66(3)\left[-6\right]$\\
& & & $m_e$ & $-2.855\left[-4\right]$ & $-8.832\left[-9\right]$ & $-6.126\left[-9\right]$ & $-1.17(1)\left[-7\right]$ & $\hphantom{-}5.43(5)\left[-8\right]$ & $-7.8(1)\left[-8\right]$\\
\rule[-3mm]{0mm}{4mm}
& & & $\hbar/R_N c$ & $-2.321\left[-8\right]$ & $-3.671\left[-12\right]$ & $-3.001\left[-12\right]$ & $-1.825(5)\left[-11\right]$ & $\hphantom{-}1.077(2)\left[-10\right]$ & $\hphantom{-}8.28(2)\left[-11\right]$\\

14 & 28 & 3.1224 & $0$ & $-1.027\left[-1\right]$ & $-1.790\left[-6\right]$ & $-6.073\left[-7\right]$ & $-2.51(1)\left[-5\right]$ & $\hphantom{-}6.90(1)\left[-6\right]$ & $-2.06(1)\left[-5\right]$\\
& & & $\alpha m_e$ & $-9.578\left[-2\right]$ & $-1.786\left[-6\right]$ & $-6.065\left[-7\right]$ & $-2.50(1)\left[-5\right]$ & $\hphantom{-}8.85(1)\left[-6\right]$ & $-1.85(1)\left[-5\right]$\\
& & & $Z\alpha m_e$ & $-4.584\left[-2\right]$ & $-1.447\left[-6\right]$ & $-5.337\left[-7\right]$ & $-1.98(1)\left[-5\right]$ & $\hphantom{-}7.29(1)\left[-6\right]$ & $-1.45(1)\left[-5\right]$\\
& & & $m_e$ & $-3.012\left[-3\right]$ & $-2.784\left[-7\right]$ & $-1.563\left[-7\right]$ & $-3.0(2)\left[-6\right]$ & $\hphantom{-}1.214(1)\left[-6\right]$ & $-2.2(2)\left[-6\right]$\\
\rule[-3mm]{0mm}{4mm}
& & & $\hbar/R_N c$ & $-4.962\left[-7\right]$ & $-2.380\left[-10\right]$ & $-1.681\left[-10\right]$ & $-9.44(2)\left[-10\right]$ & $\hphantom{-}1.206(1)\left[-9\right]$ & $-1.42(2)\left[-10\right]$\\

20 & 40 & 3.4776 & $0$ & $-1.475\left[-1\right]$ & $-5.143\left[-6\right]$ & $-1.739\left[-6\right]$ & $-5.94(1)\left[-5\right]$ & $\hphantom{-}1.902(2)\left[-5\right]$ & $-4.73(1)\left[-5\right]$\\
& & & $\alpha m_e$ & $-1.405\left[-1\right]$ & $-5.137\left[-6\right]$ & $-1.738\left[-6\right]$ & $-5.93(2)\left[-5\right]$ & $\hphantom{-}2.089(2)\left[-5\right]$ & $-4.53(2)\left[-5\right]$\\
& & & $Z\alpha m_e$ & $-6.614\left[-2\right]$ & $-4.159\left[-6\right]$ & $-1.525\left[-6\right]$ & $-4.68(1)\left[-5\right]$ & $\hphantom{-}1.701(2)\left[-5\right]$ & $-3.55(1)\left[-5\right]$\\
& & & $m_e$ & $-7.775\left[-3\right]$ & $-1.171\left[-6\right]$ & $-5.930\left[-7\right]$ & $-1.093(2)\left[-5\right]$ & $\hphantom{-}4.213(1)\left[-6\right]$ & $-8.48(2)\left[-6\right]$\\
\rule[-3mm]{0mm}{4mm}
& & & $\hbar/R_N c$ & $-1.906\left[-6\right]$ & $-1.511\left[-9\right]$ & $-9.812\left[-10\right]$ & $-4.23(2)\left[-9\right]$ & $\hphantom{-}5.937(5)\left[-9\right]$ & $-7.9(2)\left[-10\right]$\\

50 & 120 & 4.6519 & $0$ & $-3.919\left[-1\right]$ & $-9.165\left[-5\right]$ & $-2.830\left[-5\right]$ & $-5.3(2)\left[-4\right]$ & $\hphantom{-}1.750(1)\left[-4\right]$ & $-4.7(2)\left[-4\right]$\\
& & & $\alpha m_e$ & $-3.847\left[-1\right]$ & $-9.165\left[-5\right]$ & $-2.830\left[-5\right]$ & $-5.3(1)\left[-4\right]$ & $\hphantom{-}1.771(1)\left[-4\right]$ & $-4.7(1)\left[-4\right]$\\
& & & $Z\alpha m_e$ & $-1.842\left[-1\right]$ & $-7.571\left[-5\right]$ & $-2.481\left[-5\right]$ & $-4.2(1)\left[-4\right]$ & $\hphantom{-}1.426(1)\left[-4\right]$ & $-3.8(1)\left[-4\right]$\\
& & & $m_e$ & $-7.856\left[-2\right]$ & $-4.885\left[-5\right]$ & $-1.773\left[-5\right]$ & $-2.5(1)\left[-4\right]$ & $\hphantom{-}8.513(1)\left[-5\right]$ & $-2.3(1)\left[-4\right]$\\
\rule[-3mm]{0mm}{4mm}
& & & $\hbar/R_N c$ & $-9.165\left[-5\right]$ & $-2.958\left[-7\right]$ & $-1.344\left[-7\right]$ & $-6.25(2)\left[-7\right]$ & $\hphantom{-}2.990(5)\left[-7\right]$ & $-7.56(2)\left[-7\right]$\\

92 & 238 	& 5.8337 & $0$ & $-9.059\left[-1\right]$ & $-1.262\left[-3\right]$ & $-2.744\left[-4\right]$ & $-3.268(1)\left[-3\right]$ & $\hphantom{-}9.518(1)\left[-4\right]$ & $-3.853(1)\left[-3\right]$\\
& & & $\alpha m_e$ & $-8.986\left[-1\right]$ & $-1.262\left[-3\right]$ & $-2.743\left[-4\right]$ & $-3.268(1)\left[-3\right]$ & $\hphantom{-}9.538(1)\left[-4\right]$ & $-3.851(1)\left[-3\right]$\\
& & & $Z\alpha m_e$ & $-4.966\left[-1\right]$ & $-1.112\left[-3\right]$ & $-2.487\left[-4\right]$ & $-2.783(1)\left[-3\right]$ & $\hphantom{-}8.160(1)\left[-4\right]$ & $-3.328(1)\left[-3\right]$\\
& & & $m_e$ & $-3.970\left[-1\right]$ & $-1.017\left[-3\right]$ & $-2.311\left[-4\right]$ & $-2.488(1)\left[-3\right]$ & $\hphantom{-}7.316(5)\left[-4\right]$ & $-3.005(1)\left[-3\right]$\\
& & & $\hbar/R_N c$ & $-3.964\left[-3\right]$ & $-3.422\left[-5\right]$ & $-8.887\left[-6\right]$ & $-4.49(1)\left[-4\right]$ & $\hphantom{-}2.141(1)\left[-5\right]$ & $-4.71(1)\left[-4\right]$\\
\hline \hline
\end{tabular}
\caption{Leading one-electron contribution, and radiative corrections to the hypothetical fifth-force contributions to the $1s$ energy level, for various ions and various new boson masses $m_\phi$. All corrections are given in units of $\alpha_{\mathrm{FF}}\,A\,m_e c^2$, the product of the nuclear mass number with the New Physics coupling constant and the electron rest energy. Powers of $10$ are given between square brackets. The various contributions correspond to (leading) the diagram in Fig.~\ref{fig:LeadDiagram}, and to (radiative corrections) the diagrams a, b, c, d in Fig.~\ref{fig:DiagramRadEner}.
\label{Tab:Rad_FF_1s}}
\end{center}
\end{table*}
\begin{table*}[t]
\begin{center}
\begin{tabular}{c@{\hskip 0.075in}c@{\hskip 0.075in}c|@{\hskip 0.05in}c|@{\hskip 0.1125in}l|@{\hskip 0.1125in}l@{\hskip 0.1125in}l@{\hskip 0.1125in}l@{\hskip 0.1125in}l|@{\hskip 0.05in}l}
\hline \hline
\rule[-3mm]{0mm}{8mm}
$Z$ & $A$ & $R_N$ ($\mathrm{fm}$) & $m_\phi$ & \multicolumn{1}{c|}{$E_{\mathrm{FF}\left(2s\right)}$}\hskip 0.1125in & \multicolumn{1}{c}{$E_{\mathrm{FF}\left(2s\right)}^{\mathrm{a}}$}\hskip 0.1125in & \multicolumn{1}{c}{$E_{\mathrm{FF}\left(2s\right)}^{\mathrm{b}}$}\hskip 0.1125in & \multicolumn{1}{c}{$E_{\mathrm{FF}\left(2s\right)}^{\mathrm{c}\left(\mathrm{irr}\right)}$}\hskip 0.1125in & \multicolumn{1}{c|}{$E_{\mathrm{FF}\left(2s\right)}^{\mathrm{c}\left(\mathrm{red}\right)+\mathrm{d}}$}\hskip 0.05in & \multicolumn{1}{c}{$E_{\mathrm{FF}\left(2s\right)}^{\mathrm{rad}}$}\\
\hline 
%
6 & 12 & 2.4702 & $0$ & $-1.096\left[-2\right]$ & $-1.846\left[-8\right]$ & $-6.224\left[-9\right]$ & $-3.71(1)\left[-7\right]$ & $2.90(5)\left[-5\right]$ & $\hphantom{-}2.86(5)\left[-6\right]$\\ 
& & & $\alpha m_e$ & $-6.248\left[-3\right]$ & $-1.638\left[-8\right]$ & $-6.173\left[-9\right]$ & $-3.23(1)\left[-7\right]$ & $1.70(5)\left[-5\right]$ & $\hphantom{-}1.67(5)\left[-6\right]$\\
& & & $Z\alpha m_e$ & $-2.059\left[-3\right]$ & $-9.626\left[-9\right]$ & $-5.614\left[-9\right]$ & $-1.713(5)\left[-7\right]$ & $5.73(2)\left[-5\right]$ & $\hphantom{-}5.71(2)\left[-5\right]$\\
& & & $m_e$ & $-3.565\left[-5\right]$ & $-1.038\left[-9\right]$ & $-7.812\left[-10\right]$ & $-1.305(5)\left[-8\right]$ & $1.05(5)\left[-7\right]$ & $\hphantom{-}9.0(5)\left[-8\right]$\\
\rule[-3mm]{0mm}{4mm}
& & & $\hbar/R_N c$ & $-2.906\left[-9\right]$ & $-4.545\left[-13\right]$ & $-3.771\left[-13\right]$ & $-2.15(5)\left[-12\right]$ & $8.5(5)\left[-12\right]$ & $\hphantom{-}5.5(5)\left[-12\right]$\\

14 & 28 & 3.1224 & $0$ & $-2.571\left[-2\right]$ & $-2.265\left[-7\right]$ & $-7.647\left[-8\right]$ & $-3.091(5)\left[-6\right]$ & $6.69(2)\left[-5\right]$ & $\hphantom{-}6.35(2)\left[-5\right]$\\
& & & $\alpha m_e$ & $-1.974\left[-2\right]$ & $-2.197\left[-7\right]$ & $-7.631\left[-8\right]$ & $-2.978(5)\left[-6\right]$ & $5.93(2)\left[-5\right]$ & $\hphantom{-}5.60(2)\left[-5\right]$\\
& & & $Z\alpha m_e$ & $-4.870\left[-3\right]$ & $-1.176\left[-7\right]$ & $-6.868\left[-8\right]$ & $-1.361(1)\left[-6\right]$ & $1.38(1)\left[-5\right]$ & $\hphantom{-}1.23(1)\left[-5\right]$\\
& & & $m_e$ & $-3.750\left[-4\right]$ & $-3.122\left[-8\right]$ & $-2.041\left[-8\right]$ & $-3.074(5)\left[-7\right]$ & $1.16(1)\left[-6\right]$ & $\hphantom{-}8.0(1)\left[-7\right]$\\
\rule[-3mm]{0mm}{4mm}
& & & $\hbar/R_N c$ & $-6.259\left[-8\right]$ & $-2.947\left[-11\right]$ & $-2.136\left[-11\right]$ & $-1.086(5)\left[-10\right]$ & $2.03(2)\left[-10\right]$ & $\hphantom{-}4.4(2)\left[-11\right]$\\

20 & 40 & 3.4776 & $0$ & $-3.698\left[-2\right]$ & $-6.593\left[-7\right]$ & $-2.207\left[-7\right]$ & $-7.359(5)\left[-6\right]$ & $9.622(5)\left[-5\right]$ & $\hphantom{-}8.798(5)\left[-5\right]$\\
& & & $\alpha m_e$ & $-3.065\left[-2\right]$ & $-6.488\left[-7\right]$ & $-2.205\left[-7\right]$ & $-7.212(5)\left[-6\right]$ & $8.06(5)\left[-5\right]$ & $\hphantom{-}7.25(5)\left[-5\right]$\\
& & & $Z\alpha m_e$ & $-7.081\left[-3\right]$ & $-3.434\left[-7\right]$ & $-1.978\left[-7\right]$ & $-3.172(2)\left[-6\right]$ & $2.065(1)\left[-5\right]$ & $\hphantom{-}1.694(1)\left[-5\right]$\\
& & & $m_e$ & $-9.664\left[-4\right]$ & $-1.284\left[-7\right]$ & $-7.874\left[-8\right]$ & $-1.076(1)\left[-6\right]$ & $3.117(2)\left[-6\right]$ & $\hphantom{-}1.834(2)\left[-6\right]$\\
\rule[-3mm]{0mm}{4mm}
& & & $\hbar/R_N c$ & $-2.428\left[-7\right]$ & $-1.884\left[-10\right]$ & $-1.262\left[-10\right]$ & $-5.973(5)\left[-10\right]$ & $8.55(1)\left[-10\right]$ & $-5.7(1)\left[-11\right]$\\

50 & 120 & 4.6519 & $0$ & $-9.970\left[-2\right]$ & $-1.335\left[-5\right]$ & $-3.891\left[-6\right]$ & $-7.274(1)\left[-5\right]$ & $2.516(5)\left[-4\right]$ & $\hphantom{-}1.616(5)\left[-4\right]$\\
& & & $\alpha m_e$ & $-9.281\left[-2\right]$ & $-1.332\left[-5\right]$ & $-3.890\left[-6\right]$ & $-7.25(2)\left[-5\right]$ & $2.379(5)\left[-4\right]$ & $\hphantom{-}1.482(5)\left[-4\right]$\\
& & & $Z\alpha m_e$ & $-2.139\left[-2\right]$ & $-7.530\left[-6\right]$ & $-3.493\left[-6\right]$ & $-3.1764(2)\left[-5\right]$ & $7.627(2)\left[-5\right]$ & $\hphantom{-}3.348(2)\left[-5\right]$\\
& & & $m_e$ & $-9.941\left[-3\right]$ & $-5.434\left[-6\right]$ & $-2.582\left[-6\right]$ & $-2.2778(1)\left[-5\right]$ & $3.95(1)\left[-5\right]$ & $\hphantom{-}8.7(1)\left[-6\right]$\\
\rule[-3mm]{0mm}{4mm}
& & & $\hbar/R_N c$ & $-1.290\left[-5\right]$ & $-4.091\left[-8\right]$ & $-1.923\left[-8\right]$ & $-7.753(2)\left[-8\right]$ & $6.75(1)\left[-8\right]$ & $-7.02(1)\left[-8\right]$\\

92 & 238 	& 5.8337 & $0$ & $-2.427\left[-1\right]$ & $-2.579\left[-4\right]$ & $-4.828\left[-5\right]$ & $-5.754(2)\left[-4\right]$ & $7.009(2)\left[-4\right]$ & $-6.152(2)\left[-4\right]$\\
& & & $\alpha m_e$ & $-2.256\left[-1\right]$ & $-2.578\left[-4\right]$ & $-4.828\left[-5\right]$ & $-5.751(2)\left[-4\right]$ & $8.6(1)\left[-4\right]$ & $-4.6(1)\left[-4\right]$\\
& & & $Z\alpha m_e$ & $-7.446\left[-2\right]$ & $-1.794\left[-4\right]$ & $-4.466\left[-5\right]$ & $-3.238(1)\left[-4\right]$ & $3.72(1)\left[-4\right]$ & $-1.76(1)\left[-4\right]$\\
& & & $m_e$ & $-6.058\left[-2\right]$ & $-1.639\left[-4\right]$ & $-4.216\left[-5\right]$ & $-2.912(1)\left[-4\right]$ & $3.124(2)\left[-4\right]$ & $-1.849(2)\left[-4\right]$\\
& & & $\hbar/R_N c$ & $-6.058\left[-4\right]$ & $-6.575\left[-6\right]$ & $-1.709\left[-6\right]$ & $-8.91(2)\left[-6\right]$ & $7.21(2)\left[-6\right]$ & $-9.98(3)\left[-6\right]$\\
\hline \hline
\end{tabular}
\caption{Leading one-electron contribution, and radiative corrections to the hypothetical fifth-force contributions to the $2s$ energy level, for various ions and various new boson masses $m_\phi$. All corrections are given in units of $\alpha_{\mathrm{FF}}\,A\,m_e c^2$, the product of the nuclear mass number with the New Physics coupling constant and the electron rest energy. Powers of $10$ are given between square brackets. The various contributions correspond to (leading) the diagram in Fig.~\ref{fig:LeadDiagram}, and to (radiative corrections) the diagrams a, b, c, d in Fig.~\ref{fig:DiagramRadEner}.
\label{Tab:Rad_FF_2s}}
\end{center}
\end{table*}
\begin{table*}[t]
\begin{center}
\begin{tabular}{c@{\hskip 0.075in}c@{\hskip 0.075in}c|@{\hskip 0.05in}c|@{\hskip 0.1125in}l|@{\hskip 0.1125in}l@{\hskip 0.1125in}l@{\hskip 0.1125in}l@{\hskip 0.1125in}l|@{\hskip 0.05in}l}
\hline \hline
\rule[-3mm]{0mm}{8mm}
$Z$ & $A$ & $R_N$ ($\mathrm{fm}$) & $m_\phi$ & \multicolumn{1}{c|}{$E_{\mathrm{FF}\left(2p_{1/2}\right)}$}\hskip 0.1125in & \multicolumn{1}{c}{$E_{\mathrm{FF}\left(2p_{1/2}\right)}^{\mathrm{a}}$}\hskip 0.1125in & \multicolumn{1}{c}{$E_{\mathrm{FF}\left(2p_{1/2}\right)}^{\mathrm{b}}$}\hskip 0.1125in & \multicolumn{1}{c}{$E_{\mathrm{FF}\left(2p_{1/2}\right)}^{\mathrm{c}\left(\mathrm{irr}\right)}$}\hskip 0.1125in & \multicolumn{1}{c|}{$E_{\mathrm{FF}\left(2p_{1/2}\right)}^{\mathrm{c}\left(\mathrm{red}\right)+\mathrm{d}}$}\hskip 0.05in & \multicolumn{1}{c}{$E_{\mathrm{FF}\left(2p_{1/2}\right)}^{\mathrm{rad}}$}\\
\hline 
%
6 & 12 & 2.4702 & $0$ & $-1.096\left[-2\right]$ & $-5.172\left[-12\right]$ & $-1.036\left[-12\right]$ & $-5.101\left[-5\right]$ & $\hphantom{-}1.840\left[-5\right]$ & $-3.265\left[-5\right]$\\ 
& & & $\alpha m_e$ & $-5.920\left[-3\right]$ & $-4.043\left[-12\right]$ & $\hphantom{-}1.214\left[-11\right]$ & $-4.328\left[-5\right]$ & $\hphantom{-}8.585(2)\left[-6\right]$ & $-3.479\left[-5\right]$\\
& & & $Z\alpha m_e$ & $-6.880\left[-4\right]$ & $-1.725\left[-12\right]$ & $\hphantom{-}5.848\left[-11\right]$ & $-9.598\left[-6\right]$ & $\hphantom{-}2.50(1)\left[-7\right]$ & $-9.348(1)\left[-6\right]$\\
& & & $m_e$ & $-4.848\left[-8\right]$ & $-4.746\left[-14\right]$ & $\hphantom{-}1.789\left[-12\right]$ & $-1.027\left[-9\right]$ & $-1.158\left[-8\right]$ & $-1.261\left[-8\right]$\\
\rule[-3mm]{0mm}{4mm}
& & & $\hbar/R_N c$ & $-1.047\left[-12\right]$ & $-2.766\left[-17\right]$ & $\hphantom{-}1.210\left[-16\right]$ & $-1.912\left[-14\right]$ & $-7.210(2)\left[-14\right]$ & $-9.113(2)\left[-14\right]$\\

14 & 28 & 3.1224 & $0$ & $-2.571\left[-2\right]$ & $-3.606\left[-10\right]$ & $-7.177\left[-11\right]$ & $-1.201(1)\left[-4\right]$ & $\hphantom{-}4.250(2)\left[-5\right]$ & $-7.76(1)\left[-5\right]$\\
& & & $\alpha m_e$ & $-1.954\left[-2\right]$ & $-3.333\left[-10\right]$ & $-1.817\left[-11\right]$ & $-1.154\left[-4\right]$ & $\hphantom{-}2.966(2)\left[-5\right]$ & $-8.574(2)\left[-5\right]$\\
& & & $Z\alpha m_e$ & $-1.646\left[-3\right]$ & $-1.214\left[-10\right]$ & $\hphantom{-}7.050\left[-10\right]$ & $-2.293\left[-5\right]$ & $\hphantom{-}8.5(1)\left[-7\right]$ & $-2.208(1)\left[-5\right]$\\
& & & $m_e$ & $-2.889\left[-6\right]$ & $-9.167\left[-12\right]$ & $\hphantom{-}9.483\left[-11\right]$ & $-5.821\left[-8\right]$ & $-2.445(2)\left[-8\right]$ & $-8.276(2)\left[-8\right]$\\
\rule[-3mm]{0mm}{4mm}
& & & $\hbar/R_N c$ & $-1.236\left[-10\right]$ & $-1.178\left[-14\right]$ & $\hphantom{-}6.922\left[-15\right]$ & $-2.161\left[-12\right]$ & $-3.221(2)\left[-12\right]$ & $-5.387(2)\left[-12\right]$\\

20 & 40 & 3.4776 & $0$ & $-3.698\left[-2\right]$ & $-2.205\left[-9\right]$ & $-4.342\left[-10\right]$ & $-1.729\left[-4\right]$ & $\hphantom{-}5.74(1)\left[-5\right]$ & $-1.155\left[-4\right]$\\
& & & $\alpha m_e$ & $-3.050\left[-2\right]$ & $-2.110\left[-9\right]$ & $-3.443\left[-10\right]$ & $-1.694\left[-4\right]$ & $\hphantom{-}4.45(1)\left[-5\right]$ & $-1.249(1)\left[-4\right]$\\
& & & $Z\alpha m_e$ & $-2.428\left[-3\right]$ & $-7.536\left[-10\right]$ & $\hphantom{-}1.909\left[-9\right]$ & $-3.364\left[-5\right]$ & $-5.6(2)\left[-7\right]$ & $-3.420(2)\left[-5\right]$\\
& & & $m_e$ & $-1.562\left[-5\right]$ & $-9.040\left[-11\right]$ & $\hphantom{-}4.579\left[-10\right]$ & $-3.023\left[-7\right]$ & $-1.344(2)\left[-7\right]$ & $-4.362\left[-7\right]$\\
\rule[-3mm]{0mm}{4mm}
& & & $\hbar/R_N c$ & $-9.858\left[-10\right]$ & $-1.674\left[-13\right]$ & $-3.076\left[-16\right]$ & $-1.638(1)\left[-11\right]$ & $-1.924(2)\left[-11\right]$ & $-4.364(2)\left[-11\right]$\\

50 & 120 & 4.6519 & $0$ & $-9.970\left[-2\right]$ & $-3.129\left[-7\right]$ & $-5.510\left[-8\right]$ & $-4.530(1)\left[-4\right]$ & $\hphantom{-}9.53(5)\left[-5\right]$ & $-3.586(5)\left[-4\right]$\\
& & & $\alpha m_e$ & $-9.274\left[-2\right]$ & $-3.106\left[-7\right]$ & $-5.479\left[-8\right]$ & $-4.515\left[-4\right]$ & $\hphantom{-}8.37(5)\left[-5\right]$ & $-3.687(5)\left[-4\right]$\\
& & & $Z\alpha m_e$ & $-8.412\left[-3\right]$ & $-1.248\left[-7\right]$ & $-3.935\left[-9\right]$ & $-1.044\left[-4\right]$ & $-1.22(5)\left[-5\right]$ & $-1.167(5)\left[-4\right]$\\
& & & $m_e$ & $-1.124\left[-3\right]$ & $-5.224\left[-8\right]$ & $\hphantom{-}4.445\left[-9\right]$ & $-1.655\left[-5\right]$ & $-4.92(2)\left[-6\right]$ & $-2.153(2)\left[-5\right]$\\
\rule[-3mm]{0mm}{4mm}
& & & $\hbar/R_N c$ & $-3.528\left[-7\right]$ & $-2.960\left[-10\right]$ & $-1.270\left[-10\right]$ & $-2.920(1)\left[-9\right]$ & $-1.88(1)\left[-9\right]$ & $-5.22(1)\left[-9\right]$\\

92 & 238 	& 5.8337 & $0$ & $-2.427\left[-1\right]$ & $-2.341\left[-5\right]$ & $-2.934\left[-6\right]$ & $-5.898(2)\left[-4\right]$ & $\hphantom{-}4.9(2)\left[-5\right]$ & $-5.67(2)\left[-4\right]$\\
& & & $\alpha m_e$ & $-2.356\left[-1\right]$ & $-2.337\left[-5\right]$ & $-2.934\left[-6\right]$ & $-5.893(2)\left[-4\right]$ & $\hphantom{-}4.2(2)\left[-5\right]$ & $-5.74(2)\left[-4\right]$\\
& & & $Z\alpha m_e$ & $-4.064\left[-2\right]$ & $-1.315\left[-5\right]$ & $-2.175\left[-6\right]$ & $-1.462(1)\left[-4\right]$ & $-5.3(1)\left[-5\right]$ & $-2.15(1)\left[-4\right]$\\
& & & $m_e$ & $-2.473\left[-2\right]$ & $-1.086\left[-5\right]$ & $-1.880\left[-6\right]$ & $-6.99(1)\left[-5\right]$ & $-5.29(2)\left[-5\right]$ & $-1.355(2)\left[-4\right]$\\
& & & $\hbar/R_N c$ & $-8.888\left[-5\right]$ & $-2.395\left[-7\right]$ & $-6.541\left[-8\right]$ & $-1.431(2)\left[-6\right]$ & $-7.01(1)\left[-7\right]$ & $-2.437(2)\left[-6\right]$\\
\hline \hline
\end{tabular}
\caption{Leading one-electron contribution, and radiative corrections to the hypothetical fifth-force contributions to the $2p_{1/2}$ energy level, for various ions and various new boson masses $m_\phi$. All corrections are given in units of $\alpha_{\mathrm{FF}}\,A\,m_e c^2$, the product of the nuclear mass number with the New Physics coupling constant and the electron rest energy. Powers of $10$ are given between square brackets. The various contributions correspond to (leading) the diagram in Fig.~\ref{fig:LeadDiagram}, and to (radiative corrections) the diagrams a, b, c, d in Fig.~\ref{fig:DiagramRadEner}.
\label{Tab:Rad_FF_2pmin}}
\end{center}
\end{table*}

\section{Discussion and conclusion} \label{sec:DscCcl}

In the heavy boson regime, the calculated QED corrections to the hypothetical fifth force contribution to the energy levels due to interelectronic interactions via single photon exchange, are often comparable or even larger in magnitude than the leading one-electron contribution from the fifth force. As shown in Fig.~\ref{fig:SiSn} (see Tables~\ref{Tab:IEI_FF_2s}--\ref{Tab:Rad_FF_2pmin} for detailed results), this is seen both for the $2s$ ground state of Li-like ions, and for the $2p_{1/2}$ ground state of B-like ions, and occurs more markedly for lighter ions. For $Z=6$ and $Z=14$, and for boson masses $m_\phi=m_e$ and $m_\phi=\hbar/R_N c$ (with $R_N$ the nuclear radius), the fifth force contribution to the $2p_{1/2}$ energy level due to interelectronic interactions is one to three orders of magnitude larger than the leading one-electron fifth force contribution. This result can be understood in the following way: for heavy bosons, the Yukawa potential is highly localised around the nucleus, and the photon exchange diagrams where the  core electron, which is in an $s$ state, interacts with this potential (see Fig.~\ref{fig:DiagramIEIEner}), bring a large contribution, compared to the one-electron diagram, which is suppressed by the low probability density for $p$ states around the origin. This can be described as a `photon bridge', enhancing the hypothetical New Physics correction to the levels of few-electron ions. These results mean that, for multi-electronic ions, contributions from New Physics must take account of interelectronic interactions, as was done in the present work in a rigorous QED approach, or as was done in the Hartree-Fock approach in Refs.~\cite{FifthForce,NonLinXPYb,LinXPCa}. For $Z=6$, and for the boson mass $m_\phi=m_e$, the fifth force contribution to the $2p_{1/2}$ energy level due to one-loop radiative phenomena is comparable in magnitude to the one-electron fifth force contribution. The better accessibility~\cite{LiLikeBi,RehbehnCoronal} of Li-like and B-like ions to precision laser spectroscopy, compared to that of H-like ions, makes our results highly relevant to the search for NP. Our results also motivate further investigation of the $g$ factor of light Li-like and B-like ions~\cite{CakirLiB,CalciumIS,ArapoglouAcc,FifthForceG} for the search for New Physics.

\section*{Acknowledgments}

We gratefully acknowledge helpful conversations with Klaus Blaum, Halil Cakir, Zolt\'{a}n Harman, Christoph~H. Keitel, Bastian Sikora and Sven Sturm.

\appendix

\section{Derivation of the bosonic loop potential} \label{appdx:BLoop}

In this appendix we provide the derivation of the bosonic loop potential given in Eq.~(\ref{eq:BosonLoopPot}). We start by writing down the one-loop vacuum polarization (VP) correction to the photon propagator~\cite{VladZVP,Weinberg1}:
\begin{equation} \label{eq:OneLoopPhotonProp}
D_{\mu\nu}^{\left(1\right)}\left(k\right)=\left(\frac{\alpha}{\pi}\right)\eta_{\mu\nu}\int_0^1\mathrm{d}z\frac{z^2\left(1-\frac{z^2}{3}\right)}{4\left[m_e^2-\frac{1}{4}\left(1-z^2\right)k^2\right]}.
\end{equation}
Hence the bosonic loop potential (see Eq.~(\ref{eq:FFPotentialFourier}))
\begin{equation} \label{eq:BLoopFromProt}
\begin{aligned} [b]
V_{\mathrm{BL}}\left(\mathbf{k}\right)&=D_{00}^{\left(1\right)}\left(\mathbf{k}\right)V_{\mathrm{FF}}\left(\mathbf{k}\right)\\
&=-\hbar c\,4\pi\,\alpha_{\mathrm{FF}}\,A\,\frac{1}{\mathbf{k}^2+\left(\frac{m_\phi c}{\hbar}\right)^2}\left(\frac{\alpha}{\pi}\right)\\
&\times\int_0^1\mathrm{d}z\frac{z^2\left(1-\frac{z^2}{3}\right)}{4\left[m^2+\frac{1}{4}\left(1-z^2\right)\mathbf{k}^2\right]}.
\end{aligned}
\end{equation}
This potential is then Fourier-transformed to configuration space:
\begin{widetext}
\begin{equation} \label{eq:BLoopBackConfig}
\begin{aligned} [b]
V_{\mathrm{BL}}\left(\mathbf{r}\right)&=\int\frac{\mathrm{d}\mathbf{k}}{\left(2\pi\right)^3}\mathrm{e}^{\mathrm{i}\mathbf{k}\cdot\mathbf{r}}\,V_{\mathrm{BL}}\left(\mathbf{k}\right)\\
&=-\hbar c\,4\pi\,\alpha_{\mathrm{FF}}\,A\left(\frac{\alpha}{\pi}\right)\left(-\frac{\mathrm{i}}{4\left|\mathbf{r}\right|}\right)\frac{1}{\left(2\pi\right)^2}\int_0^1\mathrm{d}z\left[z^2\left(1-\frac{z^2}{3}\right)\right]\int_{-\infty}^{+\infty}\mathrm{d}k\,\mathrm{e}^{\mathrm{i}k\left|\mathbf{r}\right|}\frac{k^3}{\mathbf{k}^2+\left(\frac{m_\phi c}{\hbar}\right)^2}\frac{1}{\left[m^2+\frac{1}{4}\left(1-z^2\right)\mathbf{k}^2\right]}\\
&=-\hbar c\,4\pi\,\alpha_{\mathrm{FF}}\,A\left(\frac{\alpha}{\pi}\right)\left(-\frac{\mathrm{i}}{4\left|\mathbf{r}\right|}\right)\frac{1}{\left(2\pi\right)^2}\int_0^1\mathrm{d}z\left[z^2\left(1-\frac{z^2}{3}\right)\right]4\mathrm{i}\pi\frac{\left(4m_e^2\,\mathrm{e}^{-2\frac{m_\phi c}{\hbar}\frac{\left|\mathbf{r}\right|}{\sqrt{1-z^2}}}-m_\phi^2\left(1-z^2\right)\mathrm{e}^{-\frac{m_\phi c}{\hbar}\left|\mathbf{r}\right|}\right)}{\left(1-z^2\right)\left(4m_e^2-\left(1-z^2\right)m_\phi^2\right)}\\
&=-\frac{2}{3}\frac{\hbar c}{r}\alpha_{\mathrm{FF}}\,A\left(\frac{\alpha}{\pi}\right)\int_1^{+\infty}\frac{\mathrm{d}u}{u^2}\sqrt{u^2-1}\left(1+\frac{1}{2u^2}\right)\left[\frac{\mathrm{e}^{-2u\frac{m_e c}{\hbar}r}-\left(\frac{m_\phi}{2um_e}\right)^2\mathrm{e}^{-\frac{m_\phi c}{\hbar}r}}{1-\left(\frac{m_\phi}{2um_e}\right)^2}\right]
\end{aligned}
\end{equation}
\end{widetext}
where the change of variables $u=1/\sqrt{1-z^2}$ was used in the last step. This establishes Eq.~(\ref{eq:BosonLoopPot}).

\bibliography{Biblio}
\end{document}